\documentclass[twocolumn,runningheads]{svjour2}
\smartqed  % flush right qed marks, e.g. at end of proof
\usepackage{graphicx}
\journalname{Astrophysics and Space Science}
\begin{document}

\title{INTEGRAL/XMM views on the  MeV source GRO~J1411-64 
}
\subtitle{}

\titlerunning{INTEGRAL/XMM views ...}        % if too long for running head

\author{ Diego F. Torres   et al.
        }


\institute{Diego F. Torres \at
Instituci\'o de Recerca i Estudis Avan\c{c}ats (ICREA)
\&        Institut de Ci\`encies de l'Espai (IEEC-CSIC),
             Facultat de Ciencies,
             Universitat Aut\`onoma de Barcelona,
             Torre C5 Parell, 2a planta, 08193 Barcelona, Spain
              \email{dtorres@ieec.uab.es}
\and
Shu Zhang \at
       Laboratory for Particle Astrophysics,
Institute of High Energy Physics, Beijing 100049, China
\and
Olaf Reimer \at
W. W.
Hansen Experimental Physics Laboratory, Stanford University,
Stanford, CA 94305, USA
           \and
Xavier Barcons, Amalia Corral \at
Instituto de F\'{\i}sica de Cantabria
(CSIC-UC), E-39005 Santander, Spain             
\and
Valent\'{\i} Bosch-Ramon, Josep M. Paredes \at
Universitat de Barcelona,
Av. Diagonal 647, E08028, Barcelona, Spain
\and
Gustavo E. Romero \at
Instituto
Argentino de Radioastronomia, CC5, 1894, Villa Elisa, Argentina
\and
Jin Lu Qu \at
       Laboratory for Particle Astrophysics,
Institute of High Energy Physics, Beijing 100049, China
\and
           Werner Collmar, Volker Sch\"onfelder  \at
 Max-Planck-Institut f\"ur extraterrestrische Physik, PO Box
1603, D-85740 Garching, Germany 
\and
Yousaf Butt \at
Harvard-Smithsonian Center
for Astrophysics, 60 Garden St., Cambridge, MA 02138, USA
}

\date{Received: date / Accepted: date}

\maketitle

\begin{abstract}
The COMPTEL unidentified source GRO J 1411-64 was
observed by INTEGRAL and XMM-Newton in 2005. 
The Circinus Galaxy is the only source
detected within the 4$\sigma$ location error of GRO J1411-64, but
in here excluded as the possible counterpart. 
At soft X-rays, 22 reliable and statistically
significant sources (likelihood $> 10$) were extracted and
analyzed from XMM-Newton data. Only one of these sources, XMMU
J141255.6 -635932, is spectrally compatible with GRO~J1411-64
although the fact the soft X-ray observations do not cover the
full extent of the COMPTEL source position uncertainty make an
association hard to quantify and thus risky.
At the best location of the source, detections at hard X-rays show 
only upper limits, which, together with 
MeV results obtained by 
 COMPTEL suggest the existence of a peak
in power output located somewhere between 300-700 keV for the
so-called low state. Such a spectrum resembles those  in blazars or 
microquasars, and might suggest at work by a similar scenario.
 However, an
analysis using a microquasar model consisting on a magnetized
conical jet filled with relativistic electrons,
shows that it is hard to
comply with all observational constrains. This fact and the
non-detection at hard X-rays introduce an a-posteriori question
mark upon the physical reality of this source, what is discussed 
here.
\end{abstract}

\section{Introduction}
\label{intro}
GRO J1411-64 is the strongest variable unidentified MeV source 
located near the Galactic plane. It was dicovered by COMPTEL/CGRO 
during  1995 March-July (viewing periods 414-424), 
during which the source went on a burst event  
at MeV energies (Zhang et al. 2002). The source was detected 
at $\sim$ 7$\sigma$
in the 1-3 MeV band by combining the 7 viewing periods (VPs, the
periods of observations in CGRO), according to which the best 
location was measured 
 at  (l,b) = (311.5$^{\circ}$ ,-2.5$^{\circ}$) and the
source was referred as GRO J1411-64.  The
{\it flare duration} was several months and the rather steep 
spectral shape obtained
while the source was flaring would predict a bright, hard X-ray
source, if there is no break in the spectrum, which is explored
here. 
In what follows, we present the results of the INTEGRAL
observations of this source, as well as of XMM-Newton observation
of its best location and, followingly,  the comment  concerning 
the possible nature of this source.
\section{Observation and Data Analysis}
GRO~J1411-64 was observed by INTEGRAL  during 2004 December 30 -
2005 January 6. In total, 102 science windows (scws) were carried
out to have 210 ks of effective exposure. 
Data reduction was performed using the version 5.0 of the standard
Offline Science Analysis (OSA) software, and the spectra were
fitted with XSPEC of FTOOLS 5.3.1.

\begin{figure}
\centering
  \includegraphics{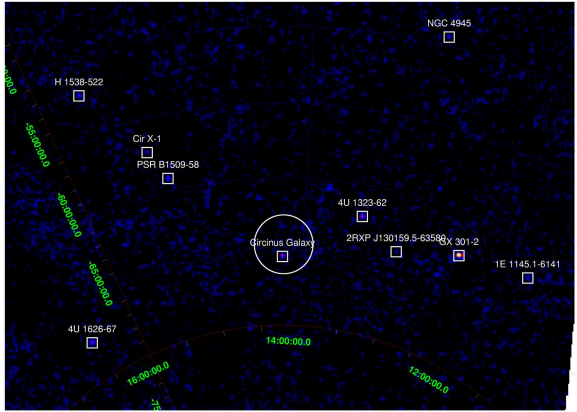}
\caption{Sky map of the GRO~J1411-64 region as seen by
IBIS/ISGRI in the 20-40 keV range, by combining all data  obtained
in the observations performed during 2004 December 30 to 2005
January 6.}
\label{fig:1}       % Give a unique label
\end{figure}

\begin{figure}
\centering
%  \includegraphics{lc_0.75-1mev.eps}
 \includegraphics{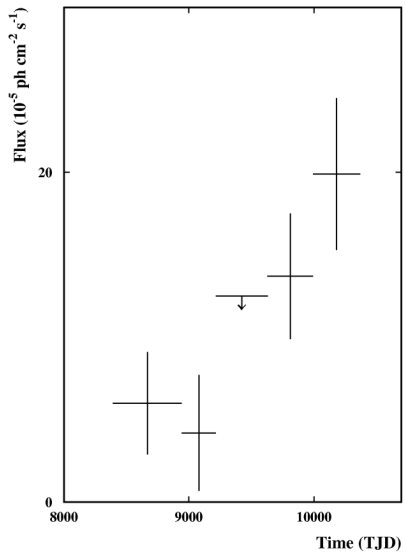}
\caption{Light curve of GRO~J1411-64 as observed by COMPTEL at
0.75-1 MeV band. The error bar is 1$\sigma$ and upper limit 2
$\sigma$. These data points include the 7 viewing periods when the
source was flaring.}
\label{fig:1}       % Give a unique label
\end{figure}
\begin{figure} 
\centering
%  \includegraphics{normal.eps}
  \includegraphics{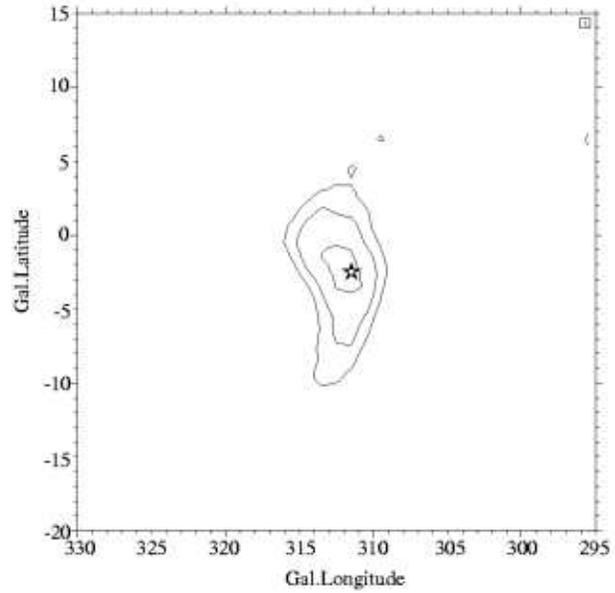} 
\caption{The skymap of GRO~J1411-64 as observed by COMPTEL in
0.75-1 MeV during 1991-1996, not including the flare period of 4
months in 1995. The star represents the best-guessed source
location. The contour lines start at a detection significance
level of 3$\sigma$ with steps of 0.5$\sigma$.}
\label{fig:1}       % Give a unique label
\end{figure}
\begin{figure}
\centering
%  \includegraphics{fig6.eps}
  \includegraphics{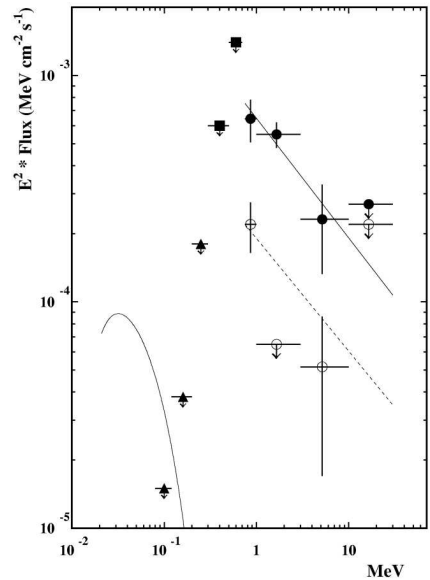}
\caption{Combined energy spectrum of GRO J1411-64. Filled (open)
circles represent flare (low) states  at MeV energies,
solid line for the flare state, dashed line for the low state,
and  the 2$\sigma$ upper limits obtained from IBIS/ISGRI
(triangles) and SPI (squares). The solid curve at low energies is
the energy spectrum of Circinus Galaxy derived from the IBIS/ISGRI
data.}
\label{fig:1}       % Give a unique label
\end{figure}

The best localization of COMPTEL source GRO J1411-64  was
observed with XMM-Newton during revolution 960 on the 7th of March
of 2005 (Obs. ID: 0204010101).  
The data were pipeline-processed with the XMM-Newton Science
Analysis Software (SAS) version 6.1.  After removal of background
flares, a total of 15.8, 15.8 and 14.6 ks of good data survived
for MOS1, MOS2 and pn respectively.
\label{sec:1}
\section{Results}
\label{sec:2}
\subsection{Hard X-rays}
\label{sec:3}

No hint of signal was found for new hard X-ray sources within the
location uncertainty of GRO~J1411-64 from individual scws of the
INTEGRAL instruments. To improve the statistics,  mosaic maps were
obtained for IBIS/ ISGRI and JEMX by combining all data. The images
of IBIS/ ISGRI were produced in the energies 20-100 keV, see Fig.
1  for the map in the 20-40 keV band as an example.
The circle holds the 4-$\sigma$  error region of GRO~J1411-64
obtained by COMPTEL during its flare in 1995 (Zhang et al. 2002).
From the possible counterparts of
GRO~J1411-64 discussed in Zhang et al.'s paper (2002), only the
Circinus Galaxy shows up in this error region as seen by INTEGRAL.
The most significant detection of Circinus Galaxy is in the energies 
20-40 keV, at a confidence level of 38$\sigma$. 
The mosaic map of JEMX  shows no significant source feature is
visible from within the 4-$\sigma$  error region.  
For SPI,  the Circinus Galaxy is at the 6$\sigma$ level in the
20-40 keV range, and it is the only source detected within the
location of GRO~J1411-64, the region of our search.

The light curve for  the Circinus Galaxy,  detected mainly by
IBIS/ISGRI, is rather constant.
The Circinus Galaxy was
investigated by Soldi et al. (2005).
Models of cutoffpl plus wabs in XSPEC can fit the data well, with
a reduced $\chi^2$ of 1.1 (7 dof). The resulting parameters are
 consistent with those in Soldi et al. 

\begin{figure}
\centering
%  \includegraphics{XMMimage.ps}
  \includegraphics[width=7cm,height=6cm]{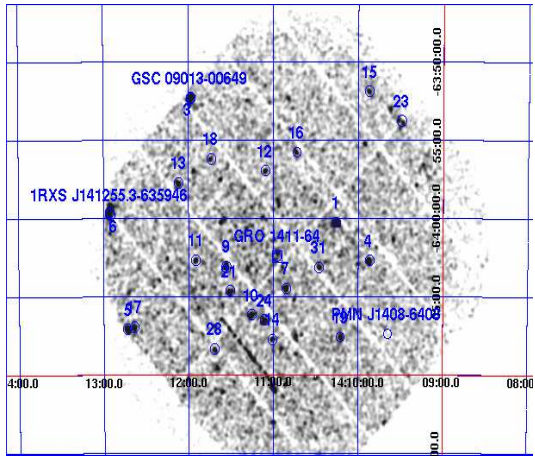}
\caption{XMM-Newton EPIC image (combining all 3 cameras), where
detected sources and previously catalogued sources have been
labeled. The centroid of the COMPTEL source GRO 1411-64 is marked
with a square box, the error contour being larger than the image
itself. See details in Torres et al. (2006).}
\label{fig:1}       % Give a unique label
\end{figure}
\begin{figure}
\centering
%  \includegraphics{XMMspectrum1.eps}
  \includegraphics{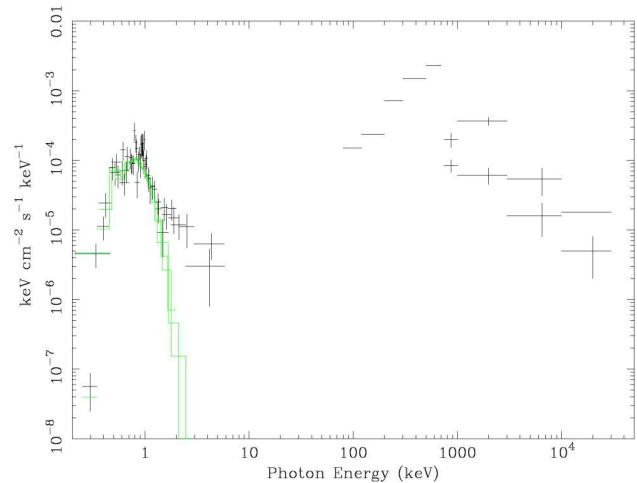}
\caption{XMM-Newton unfolded spectrum of the X-ray source XMMU
J141255.6-635932. The model shown is only the thermal component in
the X-ray spectrum. The COMPTEL detections and the INTEGRAL upper
limits are also shown at high energies, with horizontal bars
denoting 2$\sigma$ upper limits.}
\label{fig:1}       % Give a unique label
\end{figure}
\begin{figure}
\centering
%  \includegraphics{hecomptel.eps}
  \includegraphics{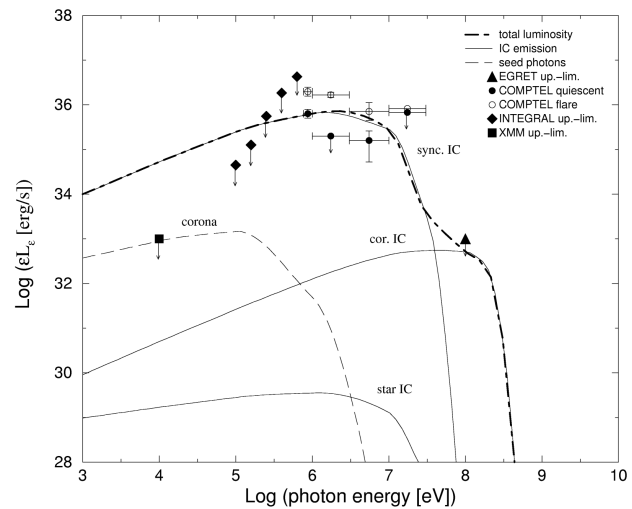}
\caption{A microquasar model on the light of observational
constraints. See Torres et al. (2006) for details.}
\label{fig:1}       % Give a unique label
\end{figure}
GRO~J1411-64 shows likely persistent emission in 0.75-1 MeV band  
during its low state (Fig. 2), where the source was detected at
$\sim 4\sigma$ by COMPTEL (Fig.3).  The corresponding spectrum of the
low state can be represented by a power law shape with spectral
index 2.5$^{+0.6}_{-0.4}$ (see Fig. 4).
Circinus Galaxy can be safely ruled out as the counterpart due to 
its spectral extrapolation well below the ones at MeV energies. 
The ISGRI/SPI upper limits combined to spectra of both flare/low
states shows the existence of a maximum in the power output at
hard X-rays, which might remind us to consider the microblazar 
as the possible source nature.
\subsection{Soft X-rays}

A total of 31 X-ray sources were formally detected by the SAS
source detection algorithm in the EPIC data. Nine of these were
excluded due to detector defects and other artifacts, in a careful
inspection. The resulting 22 reliable and statistically
significant sources (likelihood $> 10$) are shown in
Fig. 5. Among them, the unfolded spectrum (largely
independent of the model fitted) for XMMU J141255.6-635932, along
with the best fit model, the COMPTEL
detections and the INTEGRAL upper limits is plotted in Fig.6. The hard
excess exhibited by the XMM-Newton data is apparent in that
figure, and might be suggestive of a large Compton bump that would
peak in the several $\sim 100$ keV region, fitting well with the
COMPTEL detections.  However, the fact that the XMM-Newton image
does not cover the full COMPTEL source location and the
non-detection by INTEGRAL of any reliable counterpart, would make
the assumption that XMMU J141255.6-635932 is the counterpart to
GRO J1411-64, although spectrally consistent, only tentative and
risky.

%%%%% Table 1 %%%%%%%%%%%
\begin{table}[h]
\caption{Parameter values for GRO~J1411-64. At the top of table, 
parameter values for a typical
microquasar system and jet geometry are given (Bosch-Ramon et al.
2006). We have considered different values within the range open
for the free parameters finding that it is not possible to obtain
a simple microquasar model that could fit the SED. In particular,
in Fig. 7 we show a test case with the free
parameters fixed to the values presented in this table, at
the bottom. We note that, since the computed SED in Fig. 7 is
dominated in the gamma-ray band by SSC emission, the model results
would also apply for a low mass microquasar.} 
  \begin{tabular}{cl}
  \hline\noalign{\smallskip}
Parameter &  values \\
\hline\noalign{\smallskip}
Stellar bolometric luminosity [erg~s$^{-1}$] & $10^{38}$ \\
Apex dis.  to the comp. obj. [cm] & $5\times10^7$ \\
Initial jet radius [cm] & $5\times10^6$ \\
Orbital radius [cm] & $3\times10^{12}$  \\
Viewing angle to the axis of the jet [$^{\circ}$] & $45$ \\
Jet Lorentz factor & 1.2 \\
\hline\noalign{\smallskip}
%\end{tabular}
%\begin{tabular}{cl}
\hline\noalign{\smallskip} 
Jet leptonic kinetic luminosity [erg~s$^{-1}$] & $3\times10^{35}$ \\
Maximum electron Lorentz factor (jet frame) & 5$\times10^2$ \\
Maximum magnetic field [G] & 8000 \\
Electron power-law index & 1.5 \\
Total corona luminosity [erg~s$^{-1}$] & $3\times10^{33}$ \\
  \noalign{\smallskip}\hline
  \end{tabular}
  \end{table}
%%%%% Table 1 %%%%%%%%%%%
\section{Conclusion and Summary}
\label{sec:7}

The observations, subsequent analysis and
theoretical investigations pursued  shed light upon the
nature of GRO J1411-64. The combined INTEGRAL,  XMM-Newton and 
COMPTEL observations reveal no obvious counterpart at high energies 
(hard X-rays and gamma-rays). Nevertheless,  
the unique peak of the power output at these energies 
resembles  the SED seen in microquasars,
and  suggests at work by a similar scenario. 
However, an
analysis using a microquasar model consisting on a magnetized
conical jet filled with relativistic electrons which radiate
through synchrotron and inverse Compton scattering with star,
disk, corona and synchrotron photons shows that it is hard to
comply with all observational constrains (Fig. 7). The best fit parameters 
see Table 1. This fact and the
non-detection at hard X-rays introduce an a-posteriori question
mark upon the physical reality of this source. See more details in Torres 
et al. (2006). 
GLAST observations would help improving the location of the MeV
source if radiation at higher energies is not completely
suppressed, and would open the door for more efficient
multiwavelength searches of the counterpart. However, it is true
that the nature of this COMPTEL source might not be constrained
further if this detection was a one-time only transient phenomena.
GLAST will only be able to help if a candidate counterpart is
caught in the act (flaring/quiescent state of an AGN or a more
rare galactic object). Having at hand GLAST observations, in any
case, will make our currently reported investigation to naturally
fit into the testing of any hypothesis on the nature of GRO
J1411-64.

%%
%%% For two-column wide figures use
%\begin{figure*}
%\centering
%  \includegraphics[width=0.75\textwidth]{fig1.eps}
%\end{figure*}
\begin{acknowledgements}
We thank Dr. M.T. Ceballos for her help with the XMM-Newton data.
DFT has
been supported by Ministerio de Educaci\'on y Ciencia (Spain)
under grant AYA-2006-0530, and the Guggenheim Foundation. S. Zhang was subsidized by the Special Funds
for Major State Basic Research Projects and by the National
Natural Science Foundation of China. XB and AC were financially
supported for this research by the Ministerio de Educaci\'on y
Ciencia (Spain), under project ESP2003-00812. VB-R and JMP have
been supported by Ministerio de Educaci\'on y Ciencia (Spain)
under grant AYA-2004-07171-C02-01, as well as additional support
from the European Regional Development Fund (ERDF/FEDER). VB-R has
been additionally supported by the DGI of the Ministerio de
(Spain) under the fellowship BES-2002-2699. GER was supported by
grants PIP 5375 y PICT 03-13291.
\end{acknowledgements}
% BibTeX users please use
%\bibliographystyle{spmpsci}


\end{document}